%% file: main.tex
\documentclass[aps,pra,showpacs,notitlepage,twocolumn,
superscriptaddress,nofootinbib,preprintnumbers, reprint, floatfix]{revtex4-2}
\pdfoutput=1
\usepackage{bbm}
\usepackage{mathrsfs}
\usepackage[english]{babel}
\usepackage{braket}
\usepackage{float}
\usepackage{tikz}
\usetikzlibrary{quantikz}
\usepackage{array}
\usepackage{graphicx}
\usepackage{dcolumn}
\usepackage{bm}
\usepackage{hyperref}
\usepackage{siunitx}
\usepackage{xfrac}
\usepackage[nameinlink, capitalize]{cleveref}
\usepackage{xcolor}

\definecolor{mbbg}{rgb}{1,1,0.8}
\definecolor{mbfg}{rgb}{0.7,0.7,0.5}
\definecolor{t1}{rgb}{1,0.4,0.3}
\definecolor{t2}{rgb}{0.3,0.4,1}
\definecolor{t3}{rgb}{1,.3,0.8}

\newcommand{\mbopsmall}{
	\begin{tikzpicture}[scale=0.65, baseline=-0.3em]
		\draw[fill=mbbg] (0cm,0cm) circle(0.25cm);
		\foreach \x in {90, 210, 330} {
                \draw[mbfg] (0cm,0cm) -- (\x:0.25cm);
                \filldraw[black] (\x:0.25cm) circle(0.5pt);
        }
        \node at (0cm,0cm) {\small?};
	\end{tikzpicture}
}

\def\bea{\begin{eqnarray}}
\def\eea{\end{eqnarray}}
\def\>{\rangle}
\def\<{\langle}

\begin{document}


\title{Experimental quantum channel discrimination \\ using metastable states of a trapped ion}

\author{Kyle DeBry}
\email{debry@mit.edu}
\affiliation{Department of Physics, Center for Ultracold Atoms, and Research Laboratory of Electronics \\ Massachusetts Institute of Technology, Cambridge, Massachusetts 02139, USA}
\affiliation{Lincoln Laboratory, Massachusetts Institute of Technology, Lexington, Massachusetts 02421, USA}
 
\author{Jasmine Sinanan-Singh}
\affiliation{Department of Physics, Center for Ultracold Atoms, and Research Laboratory of Electronics \\ Massachusetts Institute of Technology, Cambridge, Massachusetts 02139, USA}

\author{Colin D. Bruzewicz}
\affiliation{Lincoln Laboratory, Massachusetts Institute of Technology, Lexington, Massachusetts 02421, USA}

\author{David Reens}
\affiliation{Lincoln Laboratory, Massachusetts Institute of Technology, Lexington, Massachusetts 02421, USA}

\author{May E. Kim}
\affiliation{Lincoln Laboratory, Massachusetts Institute of Technology, Lexington, Massachusetts 02421, USA}

\author{Matthew P. Roychowdhury}
\affiliation{Lincoln Laboratory, Massachusetts Institute of Technology, Lexington, Massachusetts 02421, USA}

\author{Robert McConnell}
\affiliation{Lincoln Laboratory, Massachusetts Institute of Technology, Lexington, Massachusetts 02421, USA}

\author{Isaac L. Chuang}
\affiliation{Department of Physics, Center for Ultracold Atoms, and Research Laboratory of Electronics \\ Massachusetts Institute of Technology, Cambridge, Massachusetts 02139, USA}

\author{John Chiaverini}
\affiliation{Lincoln Laboratory, Massachusetts Institute of Technology, Lexington, Massachusetts 02421, USA}
\affiliation{Massachusetts Institute of Technology, Cambridge, Massachusetts 02139, USA}

\date{May 31, 2023}

\begin{abstract}
We present experimental demonstrations of accurate and unambiguous single-shot discrimination between three quantum channels using a single trapped $^{40}\text{Ca}^{+}$ ion. The three channels cannot be distinguished unambiguously using repeated single channel queries, the natural classical analogue.  We develop techniques for using the 6-dimensional $\text{D}_{5/2}$ state space for quantum information processing, and we implement protocols to discriminate quantum channel analogues of phase shift keying and amplitude shift keying data encodings used in classical radio communication. The demonstrations achieve discrimination accuracy exceeding $99\%$ in each case, limited entirely by known experimental imperfections.

\end{abstract}
\maketitle


\begin{figure}[htbp]
    \centering
    \includegraphics[width=0.48\textwidth]{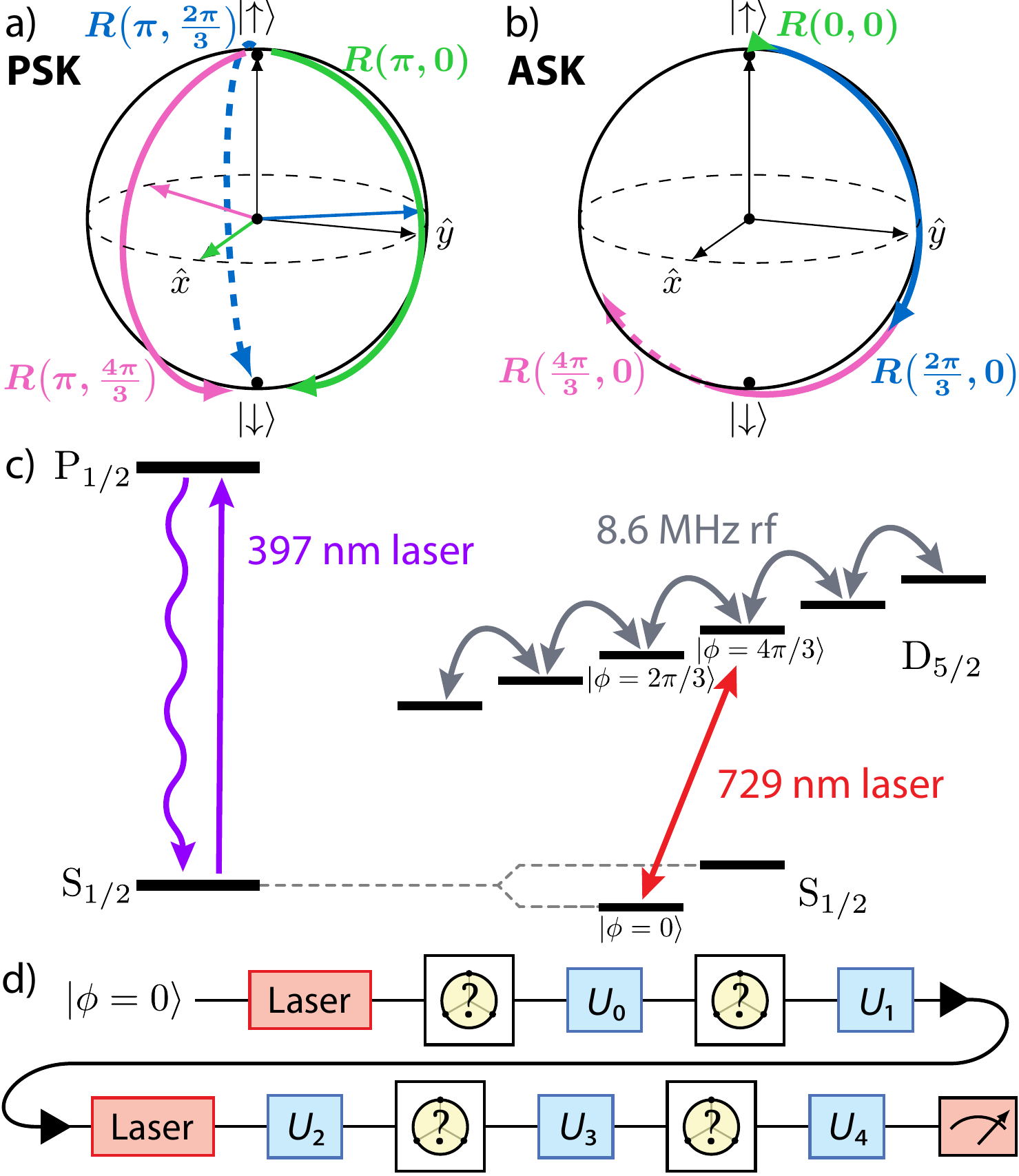}

    \caption{Operations and atomic states used in the channel-discrimination protocols. (a) Bloch sphere showing PSK rotations, with colored arrows showing rotation axes at angles of $0$, $2\pi/3$, and $4\pi/3$ in the $\hat{x}$--$\hat{y}$ plane, relative to $\hat{x}$. (b) Bloch sphere showing ASK rotations about $\hat{x}$. (c) Energy level diagram of the relevant states and transitions of $^{40}\text{Ca}^{+}$ ions. (d) Quantum circuit diagram for the PSK protocol. Gates shaded red are performed using the \SI{729}{\nano m} laser, and all others are performed with the \SI{8.6}{\mega\hertz} rf drive. Precomputed processing gates $U_i$ are shaded blue. Gates labeled with the $\protect\mbopsmall$ icon represent the oracle. 
    }
    \label{fig:energy_levels}
\end{figure}

The indistinguishability of non-orthogonal states is one of the hallmarks of quantum mechanics, and it is both an obstacle and a resource. Much theoretical and experimental effort has been devoted to the task of quantum state discrimination \cite{helstrom_quantum_1976, cook_optical_2007, chefles_quantum_2000, barnett_experimental_1997, ban1997optimum, clarke_experimental_2001, clarke_experimental_2001-1, waldherr_distinguishing_2012, solis-prosser_experimental_2017} and its applications \cite{gisin_quantum_2002, chesi_squeezing-enhanced_2018, pirandola_advances_2020} over the past several decades. The related and far richer topic of quantum \textit{channel} discrimination \cite{pirandola_fundamental_2019}
%
is significantly more complex \cite{nielsen_quantum_2000}, and many channels can be distinguished unambiguously even when analogous states cannot \cite{zhuang_ultimate_2020, rossi_quantum_2021}.
These theoretical ideas open the door to exciting experimental probes of large classes of channels, including the widely used phase-shift keying (PSK) and amplitude-shift keying (ASK) channels, which classically encode data in phase- or amplitude-modulation of a carrier signal. These protocols have natural quantum analogues where the channels cannot be distinguished without error using semiclassical finite-length protocols \cite{helstrom_quantum_1976, herzog_minimum-error_2002}.
%

\begin{figure*}[htbp]
    \centering
    \includegraphics[width=\textwidth]{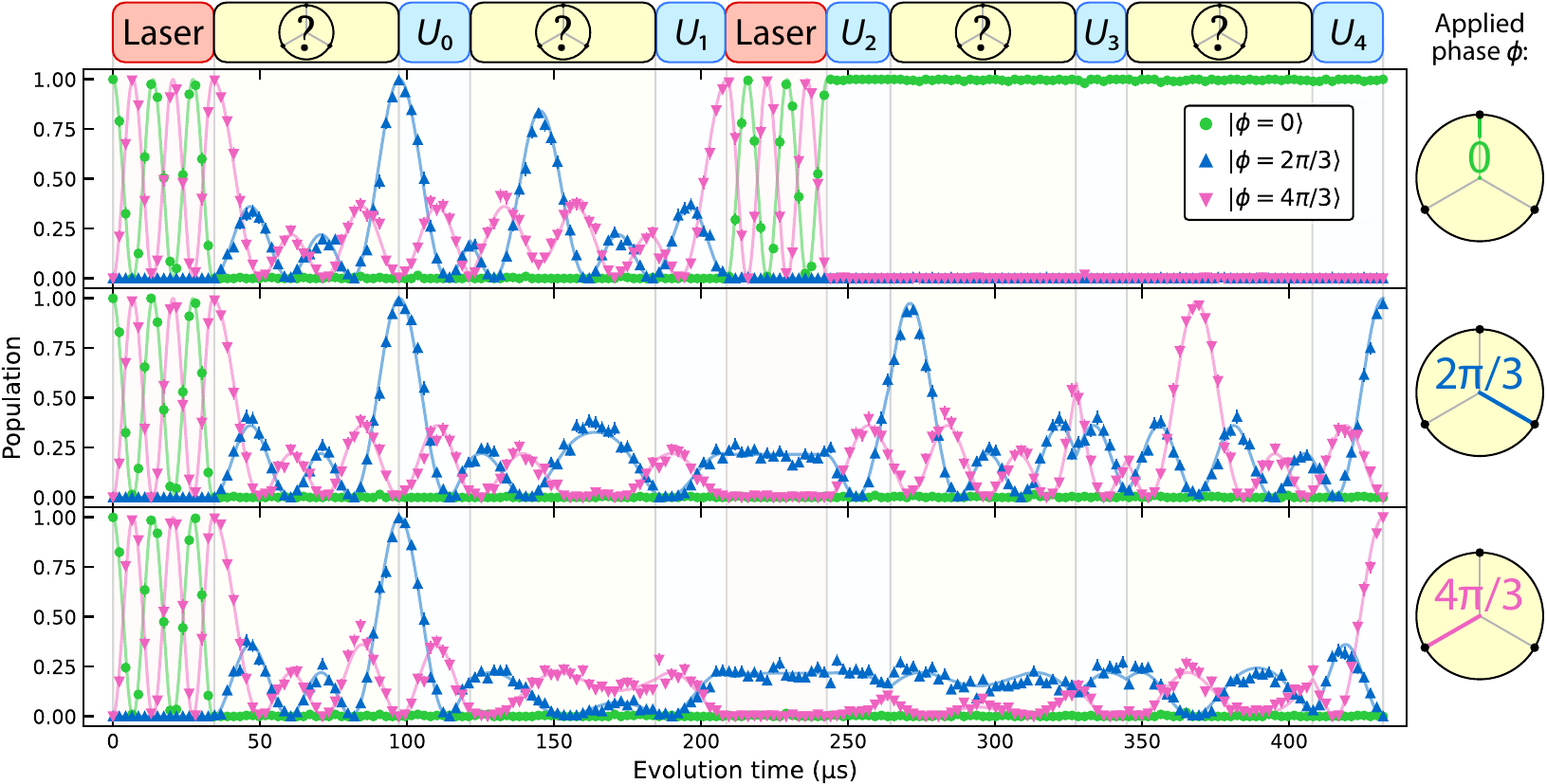}
    \caption{Experimental data of the three-phase PSK algorithm as a function of evolution time. The pulse sequence representation along the top has colors and labels matching those in \cref{fig:energy_levels}(d). The chart below is lightly shaded to highlight when each pulse is being applied. The phase of the oracle's $\pi$ pulse is indicated on the right side of the figure. The three colors and shapes of points correspond to the probability of measuring each readout state [see \cref{fig:energy_levels}(c)]. Each point corresponds to 200 trials, and the experimental data are overlaid on a zero-free-parameter simulation of the expected performance of the algorithm, displayed as solid lines. Error bars represent $1\sigma$ confidence intervals.}
    \label{fig:psk_numeric_time_series}
\end{figure*}

Distinguishing among many quantum channels requires larger Hilbert spaces and more complex quantum gate sequences than binary channel discrimination, and these needs are well-met by atomic systems. 
%
%
The long coherence times \cite{ruster_long-lived_2016, harty_high-fidelity_2014, wang_single_2021}, high-fidelity single-qubit gates \cite{harty_high-fidelity_2014, sheng2018high}, and natural presence of many long-lived states \cite{low_practical_2020} in atomic systems make them attractive for quantum protocols.  More enticingly, atoms offer high-dimensional metastable state manifolds for encoding qudits or multiple qubits within a single atom \cite{low_practical_2020, allcock_omg_2021, yang_realizing_2021, ringbauer2022universal, hrmo2023native, sun2023quantum, campbell_polyqubit_2022, gonzalez2022hardware}, which are useful for discrimination among many channels.
%
Additionally, atomic systems are well-suited for electromagnetic sensing and communication, exemplified by the elegant use of Rydberg atoms for broadband signal detection and classical PSK and ASK protocols \cite{meyer_digital_2018, simons_rydberg_2019, meyer_assessment_2020, fancher_rydberg_2021}.

In this Letter, we experimentally demonstrate solutions to quantum channel discrimination problems constructed via the formalism of quantum signal processing (QSP) \cite{low2017optimal, low2019hamiltonian, martyn_grand_2021}. QSP enables the application of nearly arbitrary $d$-degree polynomial transformations of an operator acting on a quantum subsystem by interleaving the operator with $O(d)$ unitary processing rotations. Here, we extend the protocol developed in \cite{rossi_quantum_2021} to the larger Hilbert space of the $\text{D}_{5/2}$ and $\text{S}_{1/2}$ manifolds of a trapped $^{40}\text{Ca}^{+}$ ion
and present experimental results for unambiguous channel discrimination among a triad of $\pi$-rotations about non-orthogonal axes of the Bloch sphere (see \cref{fig:energy_levels})%
; this quantum PSK scheme is a quantum channel analogue of the non-orthogonal Peres-Wootters states \cite{peres_optimal_1991}, imaginatively known as ``Mercedes-Benz'' states in classical signal processing \cite{IntroToFrames2008}. Similarly, we demonstrate and compare this with a protocol for discriminating rotations of varying angles about a consistent axis to realize a quantum ASK scheme. In both cases, we achieve detection accuracy exceeding $99\%$, with the inaccuracy well explained by known experimental imperfections. We also describe how these protocols can be extended to distinguish $n$ channels with $O(n)$ oracle queries.

{\bf Prior work.}
Quantum channel discrimination for operators from a finite set is theoretically well understood \cite{acin_statistical_2001, dariano_using_2001, duan_entanglement_2007, duan_perfect_2009, zhuang_ultimate_2020, nakahira_generalized_2021}, but there have been few experimental realizations.
The experiments reported here realize and extend the results of \cite{rossi_quantum_2021}, which give quantum algorithms with optimal query complexity that discriminate sets of quantum channels faithfully represented by a finite subgroup of SU(2). In this case, the unknown channel (the ``oracle'') is one of several possible unitary rotations.  Single-shot oracle queries can distinguish such channels only with minimum error given by the Helstrom bound \cite{herzog_minimum-error_2002, chefles_unambiguous_1998}, which is $p_\text{error} = 1/3$ for the symmetric unitary channels investigated here \cite{ban1997optimum, bae_quantum_2015, solis-prosser_experimental_2017}.
\begin{figure}[t]
    \centering
    \includegraphics[width=0.48\textwidth]{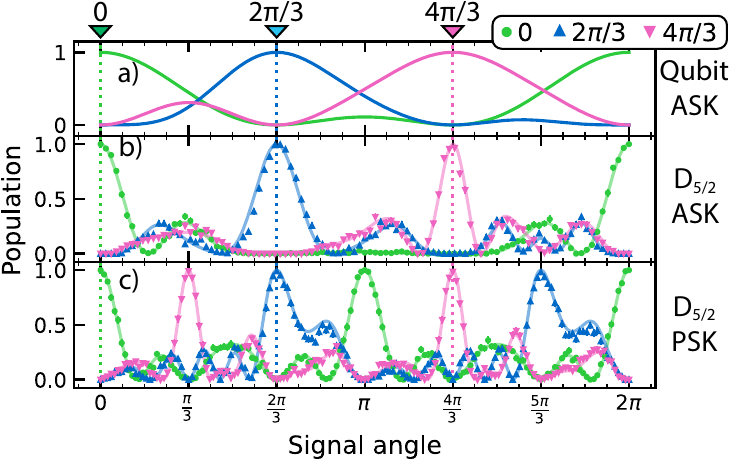}
    \caption{Input-output plot showing the response of channel discrimination protocols as a function of the signal angle for the channel. For ASK protocols, the signal angle is the rotation angle $\theta$, and for the PSK protocol it is the phase $\phi$. Error bars represent $1\sigma$ confidence intervals, but are smaller than data point symbols in most cases. Data points correspond to 200 trials. (a) Theory curves for an ASK protocol implemented with qubits [SU(2)]. (b) ASK protocol performed in the $\text{D}_{5/2}$ manifold [SU(6)], with data points overlaid on solid theory curves. (c) PSK performed in the $\text{D}_{5/2}$ manifold [SU(6)], with data points overlaid on solid theory curves.}
    \label{fig:angle_scan}
\end{figure}
\begin{figure}
    \centering
    \includegraphics[width=0.48\textwidth]{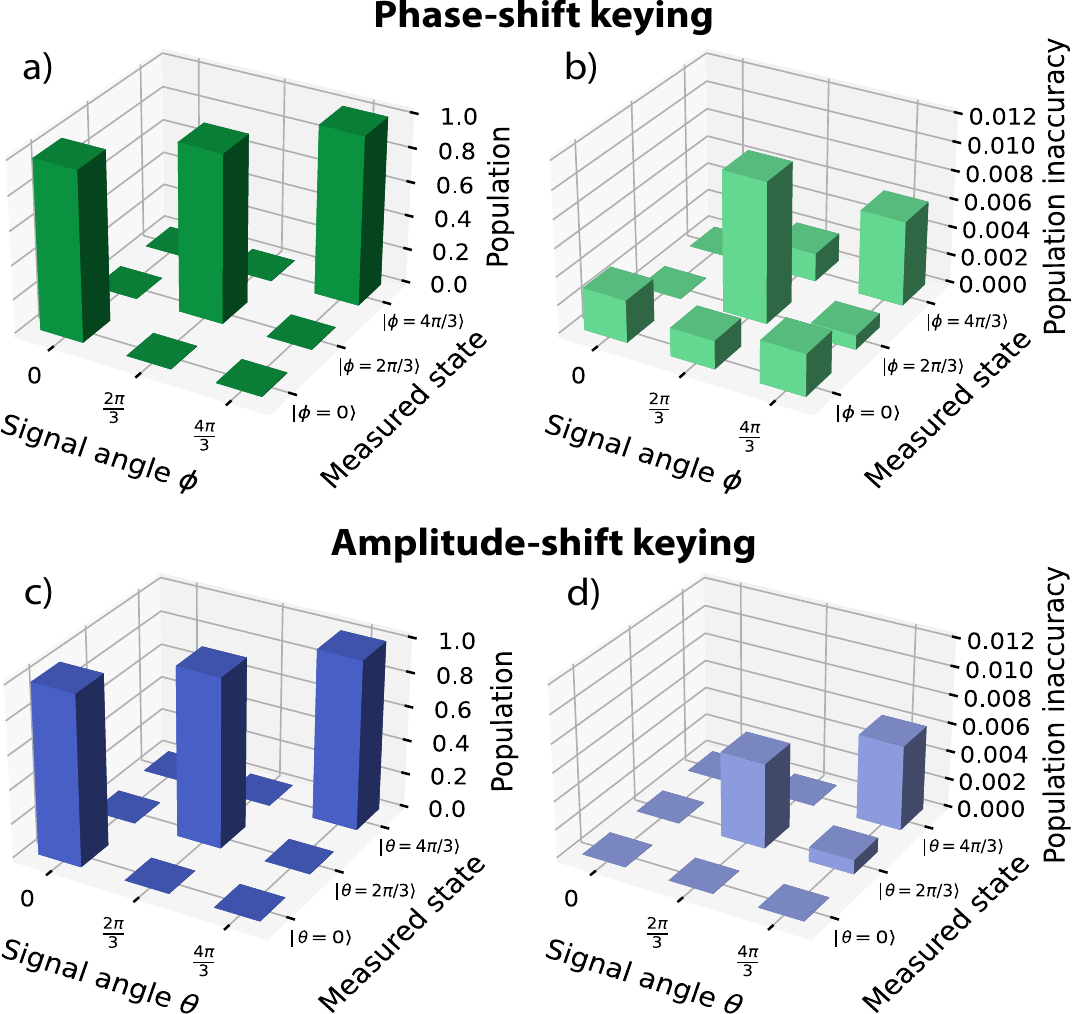}
    \caption{Populations and magnitudes of population deviations from optimal (the $3\times 3$ identity matrix) for each possible oracle value for both phase- and amplitude-shift keying implementations of the three-channel discrimination problem. The data for each signal angle corresponds to 10,000 trials. (a) PSK populations. (b) Magnitude of PSK deviation from optimal. (c) ASK populations. (d) Magnitude of ASK deviation from optimal. }
    \label{fig:bar_plot}
\end{figure}
%
Photonic systems have been used to discriminate between two such unitary channels using finite queries with \cite{laing_experimental_2009} and without \cite{zhang_linear_2008} entanglement.  Discrimination between bosonic optical channels has also been realized in the framework of quantum reading \cite{QuantumReading2021}.  Quantum process tomography has been realized in many physical systems, and such tomography can be used for channel discrimination, but aims at parameter measurement instead of making discrete decisions \cite{nielsen_quantum_2000}.Recently, another application of quantum signal processing (for Hamiltonian simulation) was implemented on a trapped-ion system \cite{kikuchi2023realization}.

{\bf Methods.}
We develop and implement pulse sequences to perform single-shot (requiring only a single trial) discrimination among three unitary channels using only four oracle queries per trial, based on the QSP-derived protocols in \cite{rossi_quantum_2021}.
In particular, the three possible unitary rotations for our implementation of PSK are the $\pi$-rotations of the Bloch sphere about different axes shown in \cref{fig:energy_levels}(a): $R(\pi, 0)$, $R(\pi, 2\pi/3)$, or $R(\pi, 4\pi/3)$, where $R(\theta, \phi)$ represents a rotation of angle $\theta$ about the axis $\left(\hat{x} \cos \phi + \hat{y} \sin \phi\right)$. This version of the algorithm was developed using numerical optimization of pulse parameters. For the case of ASK channel discrimination, the oracle becomes one of the three rotations about the $\hat{x}$ axis shown in \cref{fig:energy_levels}(b): $R(0, 0)$, $R(2\pi/3, 0)$, or $R(4\pi/3, 0)$. The ASK processing pulses were found by using \textsc{pyqsp} \cite{chao_finding_2020, martyn_grand_2021} to generate the quantum signal processing phases. In each case, the first half of the pulse sequence differentiates between signal angles (the oracle's angle $\phi$ or $\theta$ for PSK or ASK, respectively) $0$ and $\neq 0$, and the second half differentiates signal angles $2\pi/3$ and $4\pi/3$. Each half is composed of the unknown signal operator interleaved between the precomputed processing gates. This is illustrated as a circuit diagram for the PSK case in \cref{fig:energy_levels}(d). The pulse sequence for PSK (shown in a simplified form at the top of \cref{fig:psk_numeric_time_series}) is comprised of 24 laser and rf pulses, and the ASK sequence is comprised of 31 pulses. See the Supplemental Material \cite{supplemental_material} for the full pulse sequence parameters.

\nocite{bruzewicz2016scalable, roos1999quantum, sherman_experimental_2013, bazavan_synthesizing_2022, bergou2012optimal, giorda_universal_2003}

Although QSP is traditionally considered in the context of qubits (two-level systems), we have shown the ability to convert the qubit-based ASK algorithm given in \cite{rossi_quantum_2021} to both ASK and PSK algorithms in the six-level $\text{D}_{5/2}$ manifold of $^{40}\text{Ca}^{+}$ by proper consideration of the relevant SU(6) dynamics \cite{cook1979coherent, curtis2010measurement}. In particular, sequences of rotations that add up to the identity remain the identity, and sequences of rotations that add up to a bit flip operation in SU(2) likewise add up to the SU(6) generalization of a \textsc{not} gate. \Cref{fig:angle_scan}(a) and (b) show how the response of the QSP algorithm (the population in each output state as a function of the oracle's signal angle $\theta$ or $\phi$) changes when adapting the algorithm from SU(2) to SU(6). Specifically, the response of each algorithm is the same at the angles of interest ($0$, $2\pi/3$, and $4\pi/3$), but differs at other angles due to the differences in SU(2) and SU(6) dynamics. \Cref{fig:angle_scan}(c) shows the response of the PSK protocol as a function of the phase $\phi$ in the $\text{D}_{5/2}$ manifold, where we observe twice as many peaks as in the ASK protocol [\cref{fig:angle_scan}(b)]. This is because the algorithm transforms the PSK oracle with phase $\phi$ into an ASK oracle with angle $\theta=2\phi$, leading to an ambiguity when distinguishing even numbers of channels. 
In the Supplemental Material \cite{supplemental_material}, we describe in detail the relationship between SU(2) (qubit) and SU(6) ($\text{D}_{5/2}$ manifold) rotations, the transformation of ASK oracles into PSK oracles, and the resolution of this even-channel-number PSK ambiguity using a single extra oracle query.

Our experiment takes advantage of the extended Hilbert space of ground and metastable states in single $^{40}\text{Ca}^{+}$ ions. We make use of states from the $\text{D}_{5/2}$ manifold for processing and the $\text{S}_{1/2}$ manifold for shelving and readout. 
We label our three PSK algorithm readout states $\ket{\phi=0}$ in the $\text{S}_{1/2}$ manifold, and $\ket{\phi=2\pi/3}$ and $\ket{\phi=4\pi/3}$ in the $\text{D}_{5/2}$ manifold, as depicted in \cref{fig:energy_levels}(c).
Ions are confined in a cryogenic surface-electrode trap at \SI{5}{\kelvin}, similar to systems described previously \cite{sage_loading_2012, bruzewicz_dual-species_2019}. The oracle is applied as an \SI{8.6}{\mega\hertz} radio frequency (rf) signal from a small antenna located inside the coldest stage of the cryostat, approximately \SI{3}{\centi m} from the ion. In the PSK case, the oracle rotations are realized by fixed-length pulses with phases of 0, $2\pi/3$ or $4\pi/3$, the same as classical ternary PSK. The rf antenna is also used to apply the processing gates, and all rf pulses act solely on the $\text{D}_{5/2}$ manifold. Additionally, a narrow linewidth \SI{729}{\nano m} laser is used to move population between the $\text{S}_{1/2}$ manifold and the $\text{D}_{5/2}$ manifold.

Laser-based shelving to the $\text{S}_{1/2}$ manifold allows us to adapt the multi-qubit QSP sequence into a single-shot, single-ion algorithm. If the first half of the QSP sequence determines that the oracle's angle was $0$, the population will be in the $\ket{\phi=4\pi/3}$ state, and the following laser $\pi$ pulse moves population to the state $\ket{\phi=0}$ in the $\text{S}_{1/2}$ manifold (similar to ``hiding'' and ``unhiding'' pulses in \cite{barreiro_experimental_2010}). Otherwise, population is in the other states of the $\text{D}_{5/2}$ manifold after the first half of the algorithm, so the laser pulse has no effect. The protocol then proceeds to differentiate between the remaining two angles. This pulse sequence is shown schematically at the top of \cref{fig:psk_numeric_time_series}.

Through these QSP sequences of oracle queries interleaved with precomputed processing pulses, we deterministically transfer population to different states according to the oracle value. The populated state is then determined using the following qudit-style readout scheme \cite{low_practical_2020, campbell_polyqubit_2022, ringbauer2022universal}. We first apply \SI{397}{\nano\meter} detection light resonant with the $\text{S}_{1/2} \leftrightarrow \text{P}_{1/2}$ cycling transition [\cref{fig:energy_levels}(c)] and look for fluorescence, which indicates the ion was in the $\ket{\phi=0}$ state. If no fluorescence is observed, a laser $\pi$ pulse is used to transfer the population from the $\ket{\phi=2\pi/3}$ state to the ground state, and the detection beam is again applied. If there is again no fluorescence, the $\ket{\phi=4\pi/3}$ population is transferred down and the fluorescence measurement is repeated a third time. We measure the population in all three readout states to detect any leakage out of this three-state space.

{\bf Sources of error.} There are two primary sources of error for this experiment that arise from use of the larger Hilbert space of trapped ion systems: control instability and level instability. Control instability refers to amplitude or phase fluctuations of the applied pulses, and level instability refers to fluctuations in the energy of states. We minimize control instability errors by taking advantage of the superior phase and amplitude stability of the rf drive over laser pulses wherever possible by performing all the quantum information processing with the rf drive in the $\text{D}_{5/2}$ manifold. For the few remaining required laser $\pi$ pulses, we use CP Robust 180 pulse sequences \cite{ryan_robust_2010, Christensen2020}, which reduce laser-induced errors in our experiment by a factor of 5.
%
%

Use of the full manifold of states places strict requirements on the stability of the state's energy levels. The magnetic field sensitivity of Zeeman sublevels' energies passes this constraint on to the magnetic field, and we address this requirement both passively and actively. To achieve ${\tt>} 99\%$ accuracy, the length of the algorithm and our experimental parameters dictate that the rf drive must remain within \SI{30}{\hertz} of resonance, corresponding to a magnetic field stability of better than \SI{20}{\micro G}. The cryogenic apparatus allows us to stabilize the magnetic field with two rings of superconducting niobium \cite{gabrielse_superconducting_1991, wang_demonstration_2010, bruzewicz_dual-species_2019}, enabling metastable Zeeman qubit coherence times of $T_2^* \approx \SI{90}{\milli\second}$ (see Supplemental Material \cite{supplemental_material}). This stability is sufficient for a single trial taking less than \SI{1}{\milli\second}, but slow frequency drifts require compensation on the timescale of seconds, which we implement as an active feed-forward protocol. This clock-like scheme is described in the Supplemental Material \cite{supplemental_material}. This protocol brings magnetic-field-induced errors well below $1\%$, allowing us to take full advantage of these magnetic field-sensitive states.

{\bf Results.}
We achieve better than $99\%$ accuracy for both the ASK and PSK quantum channel discrimination algorithms. For the PSK case, the applied operator was correctly determined in $99.4^{+0.1}_{-0.2}\%$ of trials, averaged over the three possible oracle values. For the ASK case, we measure the correct output state with $99.6^{+0.1}_{-0.1}\%$ accuracy, even when applying a total of 31 pulses (see Supplemental Material \cite{supplemental_material}). Uncertainties represent 1$\sigma$ confidence intervals computed using the Wilson score interval method \cite{wilson_probable_1927, wilson_footnote,*brown2001interval}. The probability of detecting the ion in each of the three output states is shown in \cref{fig:bar_plot}(a) and (c), with the population deviations from the ideal case shown in \cref{fig:bar_plot}(b) and (d), for the PSK and ASK cases, respectively. A plot of the populations of the three readout states as a function of evolution time during the algorithm is shown in \cref{fig:psk_numeric_time_series} for the PSK algorithm, demonstrating close agreement of the data with the analytic predictions (solid lines). Both versions of the algorithm are shown as a function of signal angle ($
\theta$ or $\phi$) in \cref{fig:angle_scan}, again showing close agreement with predicted performance.

The inaccuracy is well-explained by known error sources, and is dominated by control errors rather than intrinsic errors. We measure the inaccuracy due to state preparation, detection, and laser $\pi$ pulses to be $0.21(5)\%$, and calculate the magnetic field-induced errors to be 0.2--0.4\% for the different versions of the algorithm by estimating the average detuning from resonance. Errors from spontaneous emission from the $\text{D}_{5/2}$ states during the duration of the experiment are ${\tt <} 0.1\%$. All three of these error mechanisms could result in population leakage out of the three-state readout space, but with the current readout method, spontaneous emission appears not as leakage but as measurement of an incorrect readout state. This is discussed in the Supplemental Material \cite{supplemental_material}.






{\bf Outlook and conclusion.}  
The results we report here set the stage for the exploration of large sets of quantum channels. As expected from theory \cite{rossi_quantum_2021}, these results show a quantum advantage over semiclassical and naïve quantum protocols (see the Supplemental Material \cite{supplemental_material}). Scaling this algorithm to $n$ channels with equally-spaced angles is straightforward and asymptotically optimal in query complexity  \cite{rossi_quantum_2021}, scaling as $O(n)$.
We have also shown how to discriminate channels in the desirable case of phase-shift keying with an even number of angles with a single additional oracle query \cite{supplemental_material}.

In conclusion, we have demonstrated entanglement-free discrimination of more than two quantum channels with finitely many queries of the operator, using a framework that can be scaled to many channels. We have demonstrated this protocol with single-shot readout by using the combined state space of the $\text{D}_{5/2}$ and $\text{S}_{1/2}$ manifolds of a trapped ion, including the development of novel quantum information processing techniques in the 6-level $\text{D}_{5/2}$ manifold that highlight the quantum information processing potential of metastable states \cite{allcock_omg_2021}. Due to atomic systems' ability to receive electromagnetic signals across an exceptionally broad frequency band, this opens the door to single-atom reception, decoding, and quantum processing of a wide range of classical and quantum signals.

\begin{acknowledgments}
We thank Z.\,M. Rossi and J.\,M. Martyn for helpful discussions. This research was supported by the U.S. Army Research Office through grant W911NF-20-1-0037. This material is based upon work supported by the National Science Foundation Graduate Research Fellowship under Grant No. 2141064. K.\,D. acknowledges support from the MIT Center for Quantum Engineering -- Laboratory for Physical Sciences Doc Bedard Fellowship. I.\,L.\,C. acknowledges support by the NSF Center for Ultracold Atoms.  The opinions, interpretations, conclusions, and recommendations are those of the authors and are not necessarily endorsed by the United States Government.
\end{acknowledgments}


\bibliography{refs}

\clearpage
\newpage
\clearpage

\input{supp_body}

\end{document}

%% file: supp_body.tex

\setcounter{section}{0}
\setcounter{equation}{0}
\setcounter{figure}{0}
\renewcommand{\thesection}{S-\Roman{section}}
\renewcommand{\thetable}{S\arabic{table}}
\renewcommand{\theequation}{S\arabic{equation}}
\renewcommand{\thefigure}{S\arabic{figure}}


\onecolumngrid

\clearpage

\begin{center}
    \textbf{ \large{Supplemental Material for: \\ Experimental quantum channel discrimination \\ using metastable states of a trapped ion}}
\end{center}

\twocolumngrid

\section{Introduction}
    This Supplemental Material contains additional details of the experimental implementation of the algorithms described in the main text, as well as additional theoretical background. \Cref{sec:experiment} expands on the description of the experimental methods, beginning with a more detailed description of the apparatus in \cref{sec:system}. We then discuss the active magnetic field feed-forward scheme in \cref{sec:magnetic}, followed by an enumeration of the pulse parameters implemented in the experiments reported in the main text in \cref{sec:pulses}. \Cref{sec:theory} contains background and additional detail for the quantum signal processing algorithms, as well as a more thorough description of the physics of rotations of the $\text{D}_{5/2}$ manifold. \Cref{sec:derivation} contains a derivation of the pulse sequences, including a review of the ideas of quantum signal processing. This is expanded on in \cref{sec:psk}, where we extend the theoretical results of QSP to allow discrimination of phase shift keying oracles. \Cref{sec:queries} discusses the query complexity of these results and the quantum advantage afforded by these QSP sequences over incoherent methods. \Cref{sec:su6} discusses the differences and similarities between the $\text{D}_{5/2}$ manifold's six-level space and QSP's typical domain of qubits, and how many QSP algorithms can be directly converted into pulse sequences in higher-spin spaces like the $\text{D}_{5/2}$ manifold. 

\section{Experimental methods}\label{sec:experiment}

\subsection{Experimental setup}\label{sec:system}
\begin{figure}
    \centering
    \includegraphics[width=0.3\textwidth]{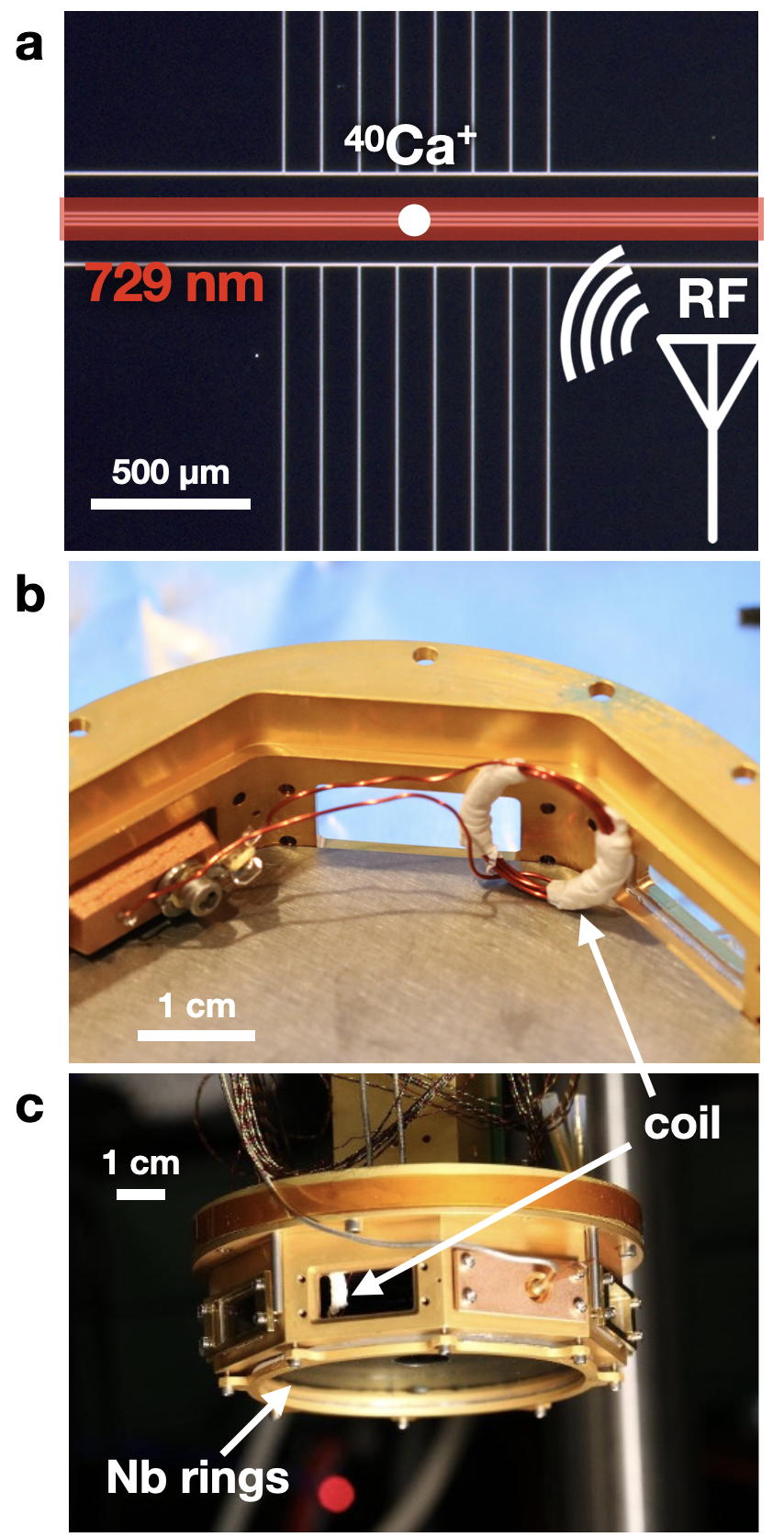}
    \caption{Experimental setup. (a) Image of the surface-electrode trap, including the orientation of the \SI{729}{\nano m} beam and a symbolic depiction of the direction of the rf antenna. (b) Image of the rf antenna, which is located within the \SI{5}{\kelvin} stage of the cryostat. (c) Image of the outside of the \SI{5}{\kelvin} stage of the cryostat, including the bottom of the two superconducting Nb rings and the rf antenna.}
    \label{fig:experiment}
\end{figure}

The experiments reported in the main text were carried out in a cryogenic surface-electrode trap \cite{sage_loading_2012}, depicted in \cref{fig:experiment}. \Cref{fig:experiment}(a) shows an optical image of the trap surface, superimposed with the beam orientation of the narrow linewidth \SI{729}{\nano m} laser that is used to drive transitions between ground and metastable states. It also depicts the approximate orientation of the rf antenna that applies processing and oracle pulses, and the antenna is shown in \cref{fig:experiment}(b). The $\approx\SI{5}{G}$ magnetic field that sets the quantization axis is oriented normal to the plane of the trap. The field is generated initially by a set of Helmholtz coils, but after cooling to below the transition temperature of niobium, the field is fixed in place by two superconducting rings of niobium and the Helmholtz coils are turned off. The bottom of the two rings is visible in \cref{fig:experiment}(c).

The system is loaded from a pre-cooled source of $^{40}\text{Ca}$ atoms provided by a two-dimensional magneto-optical trap \cite{bruzewicz2016scalable}. Atoms are ionized inside the trapping region by a two-step photoionization process. State preparation consists of Doppler cooling using the \SI{397}{nm} laser, sideband cooling using the \SI{729}{nm} laser into the motional ground state, and optical pumping into the $\ket{\phi=0}$ state. Experimental operations follow standard procedures for trapping $^{40}\text{Ca}^{+}$ in surface traps \cite{sage_loading_2012, roos1999quantum}.

The rf antenna used for the oracle and processing pulses in the algorithm is a small hand-wound coil of wire on the \SI{4}{K} stage of the cryostat. An external resonant matching circuit impedance-matches the antenna at the $\approx \SI{8.6}{MHz}$ resonance frequency of the $\text{D}_{5/2}$ manifold, set by our magnetic field strength. Due to the differing $g$-factors of the $\text{S}_{1/2}$ and $\text{D}_{5/2}$ manifolds, the rf antenna selectively drives transitions only in the $\text{D}_{5/2}$ manifold.

\subsection{Active magnetic field compensation}\label{sec:magnetic}

\begin{figure}
    \centering
    \includegraphics[width=0.48\textwidth]{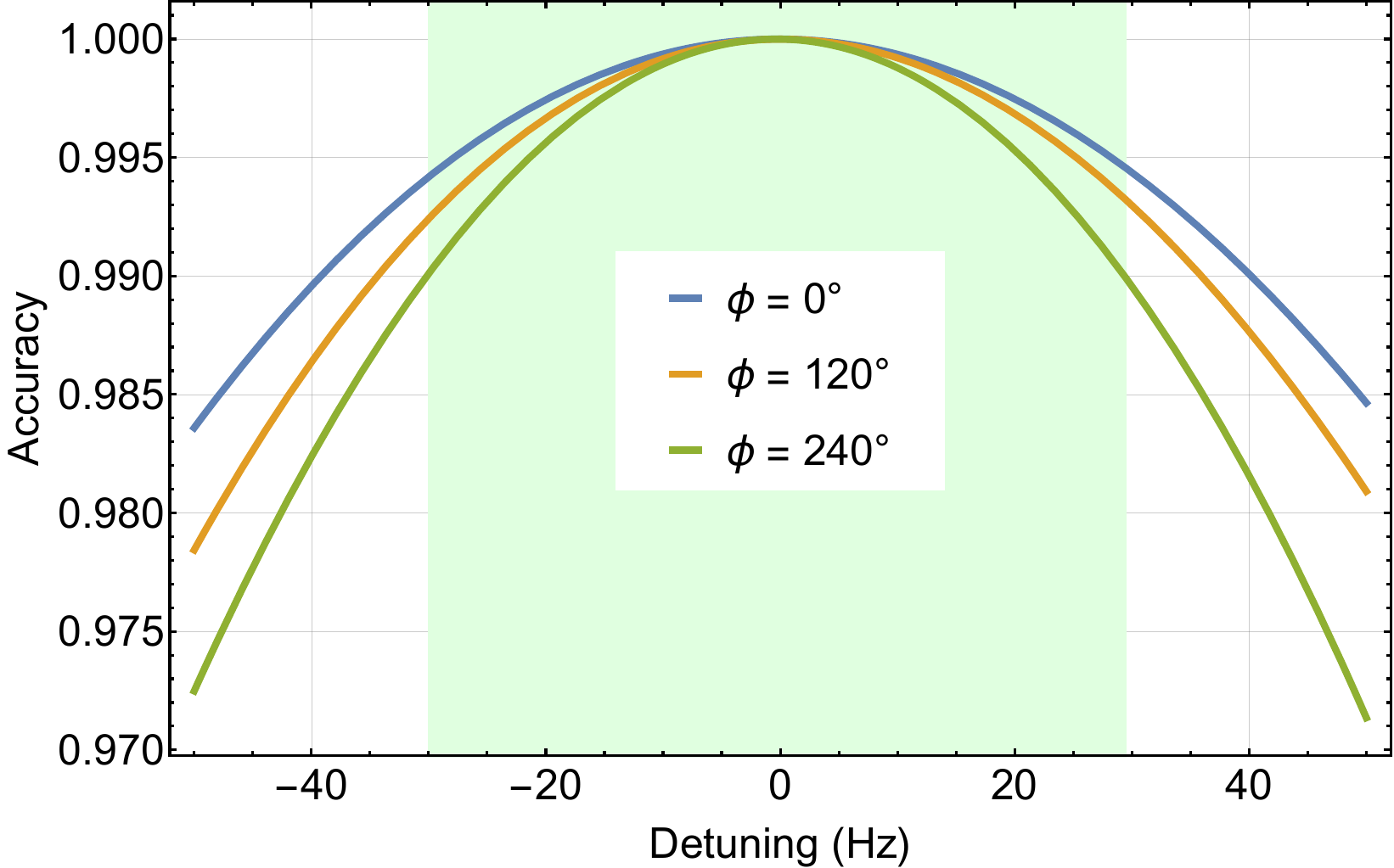}
    \caption{Analytic calculation of maximum PSK accuracy as a function of detuning of the rf drive from resonance. The green shaded region represents accuracy $\ge 99\%$ for all three oracle values.}
    \label{fig:detuning_error}
\end{figure}

As discussed in the main text, the static magnetic field seen by the ion must be kept stable to better than \SI{20}{\micro G}. By constructing propagators for the rotation operators used in the experiment that take into account the detuning of the rf source from the Zeeman resonance, we calculated the accuracy of the algorithm as a function of detuning, shown in \cref{fig:detuning_error}. This calculation assumes that the magnetic field has a constant value for each realization of the experiment, which is supported by the measured \SI{90}{\milli\second} $\text{T}_2^*$ of a metastable state Zeeman qubit (see below). While passive shielding dramatically improves the stability, we need additional compensation to reach the desired accuracy. To keep the rf drive well within \SI{30}{Hz} of resonance with the Zeeman splitting of the $\text{D}_{5/2}$ manifold, we implemented a feed-forward protocol that runs between every repetition of the experiment. We use a standard Ramsey interference clock protocol to measure the rf detuning from resonance.

\begin{figure}
    \centering
    \includegraphics[width=0.3\textwidth]{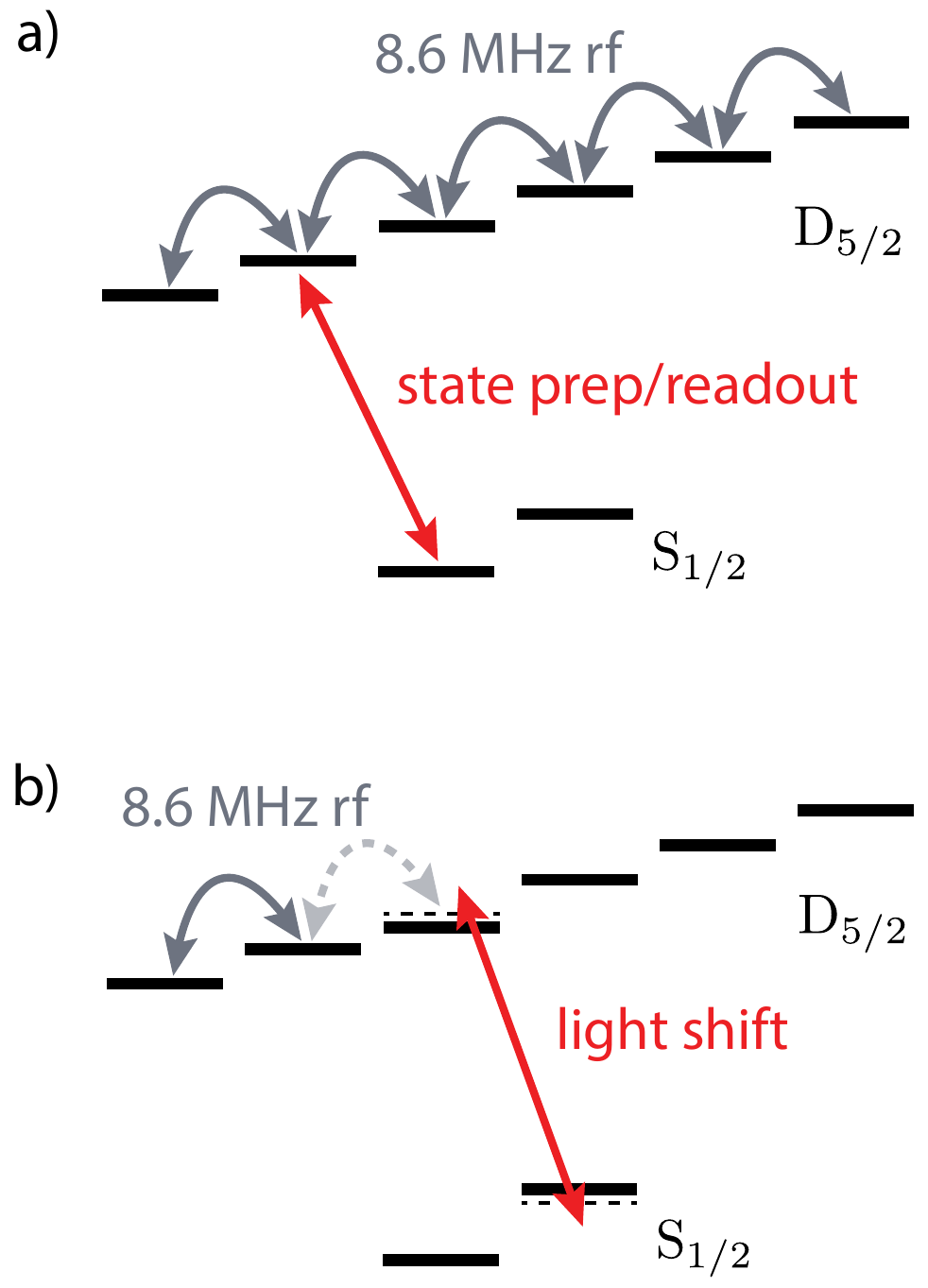}
    \caption{Energy level diagrams. (a) Energy levels showing state prep of with the \SI{729}{nm} laser into the $m=-3/2$ state in the $\text{D}_{5/2}$ manifold, and six-level coupling of the rf drive. (b) Energy levels showing the use of the \SI{729}{nm} laser to apply a light shift to the $m=-1/2$ state in the $\text{D}_{5/2}$ manifold, shifting it out of resonance with the rf drive. This isolates a qubit in the $m=-5/2$ and $m=-3/2$ states.}
    \label{fig:su6_isolation}
\end{figure}

To perform the Ramsey interference experiment, we need access to a magnetic-field-sensitive qubit. We isolate a qubit in the two lowest-energy $\text{D}_{5/2}$ states ($m=-5/2$ and $m=-3/2$ by light-shifting the neighboring state ($m=-1/2$) out of resonance with the rf drive so the drive couples the $m=-5/2$ and $m=-3/2$ states, but population is not transferred from the $m=-3/2$ state to the $m=-1/2$ state and out of our qubit Hilbert space (similar to that demonstrated in \cite{sherman_experimental_2013, bazavan_synthesizing_2022}). This is depicted in \cref{fig:su6_isolation}(b), with state preparation to the $m=-3/2$ magnetic sublevel shown in \cref{fig:su6_isolation}(a). We perform qubit isolation by detuning the \SI{729}{\nano m} laser approximately \SI{10}{\kilo\hertz} from the transition to the $m=-1/2$ state, resulting in a shift of $\approx \SI{30}{\kilo\hertz}$ for our beam intensity [\cref{fig:su6_isolation}(b)]. This shifting is applied while we simultaneously apply the rf drive to perform a $\pi/2$-pulse to set up the Ramsey experiment, and the light shift is turned off during the interrogation time. A $-\pi/2$-pulse (implemented as a $\pi/2$-pulse with a $180^\circ$ phase shift) is then applied in the same manner, and the ion is de-shelved and measured. An experimentally-calibrated constant offset of approximately \SI{100}{\hertz} is added to the measured frequency to account for unintended light shifts of the qubit states by the \SI{729}{\nano m} laser during the Ramsey sequence's $\pi/2$-pulses, which are allowed by selection rules and thus only suppressed by the few-megahertz detuning. 

\begin{figure}
    \centering
    \includegraphics[width=0.48\textwidth]{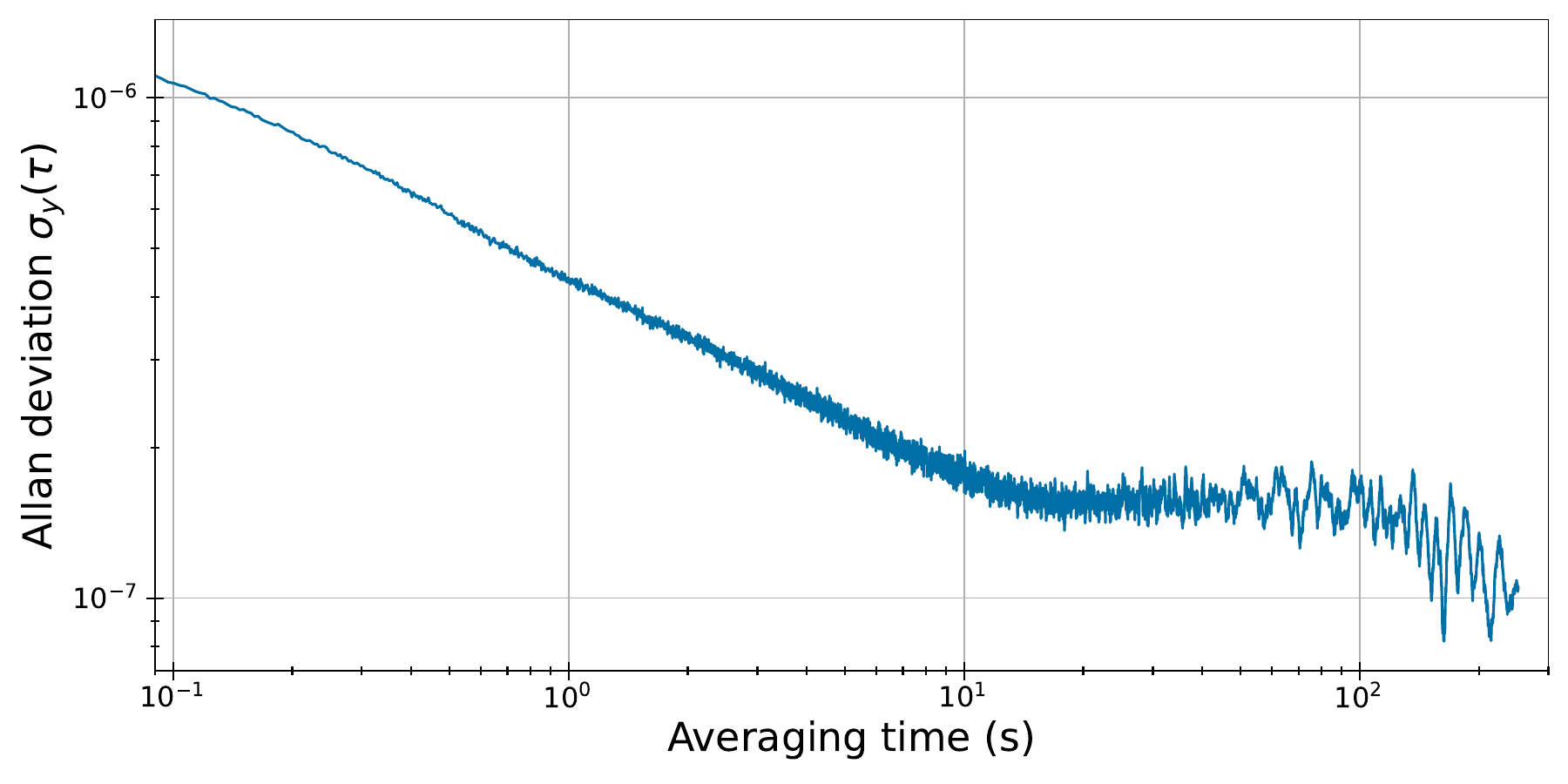}
    \caption{Allan deviation of the rf resonance frequency of the $\text{D}_{5/2}$ manifold with the superconducting niobium rings in place but no active compensation, showing drift on the timescale of tens of seconds.}
    \label{fig:allan_deviation}
\end{figure}

Slow drift of the Zeeman splitting frequency (arising from slow drifts in the static magnetic field) begins to matter on the timescale of several seconds. \Cref{fig:allan_deviation} shows the Allan deviation $\sigma_y(\tau)$ of the Zeeman qubit transition frequency for averaging time $\tau$, demonstrating the noise averaging down to $<2\times 10^{-7}$ after ten seconds, at which point slow-scale drifts begin to impact the frequency stability. Adjusting the rf frequency periodically allows us to compensate for these slow drifts. Using an interrogation time of \SI{5}{\milli\second} on either side of the fringe to determine the sign of any detuning error, and adjusting the applied rf frequency in steps of \SI{5}{\hertz}, we can bring the detuning-induced error for the full algorithm to below 1\%.

\subsection{Pulse sequences used in the experiments}\label{sec:pulses}

The following tables list the pulse sequences used in the experiments presented in the main text. In each, the pulse labeled ``Laser'' is implemented as a CP Robust 180 pulse sequence comprised of five $\pi$-pulses. Other pulses are applied by the rf antenna. A pulse with rotation and $\theta$ and phase $\phi$ is described by rotation operator $R(\theta, \phi) = e^{-i \frac{\theta}{2} \sigma_{\phi}}$, where $\sigma_\phi = \sigma_x \cos \phi + \sigma_y \sin \phi$ and $\sigma_i$ are the Pauli matrices. After this pulse sequence is applied, the readout scheme is comprised of zero, one, or two additional CP Robust 180 pulse sequences depending on the readout state. Averaging the readout pulses over the three oracle values, there are 24 total pulses applied for the PSK sequence (\cref{tab:psk_phases}), and 31 pulses applied for the ASK sequence (\cref{tab:pyqsp_phases}).

\begin{table}[]
    \centering
    \begin{tabular}{rccc}
        \# & Label & Rotation Angle $\theta$ & Phase $\phi$ \\
        \hline
        1 & Laser & $\pi$ & 0 \\
        2 & Oracle & $\pi$ & $\phi_i + \pi$ \\
        3 & $U_0$ & -1.1885 &  2.9271 \\
        4 & Oracle & $\pi$ & $\phi_i + \pi$ \\
        5 & $U_1$ & -1.1881 &  0.2146 \\
        6 & Laser & $\pi$ & 0 \\
        7 & $U_2$ & 1.0557 & -2.2241 \\
        8 & Oracle & $\pi$ & $\phi_i + \pi$ \\
        9 & $U_3$ & -0.8414 & -1.0725 \\
        10 & Oracle & $\pi$ & $\phi_i + \pi$ \\
        11 & $U_4$ & -1.1807 &  2.0282 \\
    \end{tabular}
    \caption{Pulse parameters for the phase-encoded (phase-shift keying) quantum process discrimination algorithm described in the main text. The \# column denotes the order in which the pulses are applied, and the phase $\phi$ is axis of rotation in the $\hat{x}$--$\hat{y}$ plane of the Bloch sphere (and the relative phase offset of the generated laser or rf pulse). The oracle phase $\phi_i$ can take values of $0$, $2\pi/3$, and $4\pi/3$. Laser pulses were implemented as CP Robust 180 composite pulse sequences, and couple the $\ket{\phi=0}$ and $\ket{\phi=2\pi/3}$ states.}
    \label{tab:psk_phases}
\end{table}

\begin{table}[]
    \centering
    \begin{tabular}{rccc}
        \# & Label & Rotation Angle $\theta$ & Phase $\phi$ \\
        \hline
        1 & Laser & $\pi$ & 0 \\
        2 & $U_0$ & $\pi/2$ & 0 \\
        3 & $U_1$ & 0.9603 & 0 \\
        4 & Oracle & $\theta_i$ & $\pi/2$ \\
        5 & $U_2$ & 1.2410 &  0 \\
        6 & Oracle & $\theta_i$ & $\pi/2$ \\
        7 & $U_3$ & -2.1813 &  0 \\
        8 & $U_4$ & $-\pi/2$ & 0 \\
        9 & Laser & $\pi$ & 0 \\
        10 & $U_5$ & $\pi/2$ & 0 \\
        11 & $U_6$ & 0.9603 & 0 \\
        12 & Oracle & $\theta_i$ & $\pi/2$ \\
        13 & $U_7$ & $2\pi/3$ & $\pi/2$ \\
        14 & $U_8$ & 1.2410 &  0 \\
        15 & Oracle & $\theta_i$ & $\pi/2$ \\
        16 & $U_9$ & $2\pi/3$ &  0 \\
        17 & $U_{10}$ & -2.1813 &  0 \\
        18 & $U_{11}$ & $-\pi/2$ & 0 \\
    \end{tabular}
    \caption{Pulse parameters for the angle-encoded (amplitude-shift keying) quantum process discrimination algorithm described in the main text, computed using \textsc{pyqsp}. The \# column denotes the order in which the pulses are applied, and the phase $\phi$ is axis of rotation in the $\hat{x}$--$\hat{y}$ plane of the Bloch sphere (and the relative phase offset of the generated laser or rf pulse). The oracle rotation angle $\theta_i$ can take values of $0$, $2\pi/3$, and $4\pi/3$. Laser pulses were implemented as CP Robust 180 composite pulse sequences, and couple the $\ket{\theta=0}$ and $\ket{\theta=2\pi/3}$ states.}
    \label{tab:pyqsp_phases}
\end{table}

\subsection{Discussion of error sources}\label{sec:error}

As mentioned in the main text, the dominant sources of error are errors resulting from the laser operations during state preparation and measurement, and errors resulting from detuning of the rf drive from resonance with the Zeeman splitting, which ultimately stems from magnetic field instability. The latter error is quantified as described in \cref{sec:magnetic}. We quantify the laser operation error by effectively performing the algorithm without any rf operations, i.e. we do the laser state prep, perform the mid-algorithm laser $\pi$-pulse, and perform the qudit-style readout scheme. This sequence is performed without delays, so it minimizes spontaneous decay errors. From this, we measure the $0.21(5)\%$ error reported in the main text.

The final error source is error from spontaneous decay from the $\text{D}_{5/2}$ manifold. While the other two error sources primarily result in leakage out of the three-state readout space, which could be rejected in post-selection, spontaneous decay results in the measurement of the wrong angle. (Note that we do not perform post-selection on the results, and leakage error contributes to the reported infidelities.) Spontaneous decay is also the only incoherent error source, and the only one that could not, in principle, be addressed through improved control of the system (aside from shortening the duration of the experiment). Fortunately, the experiment is significantly shorter than the \SI{1.1}{s} lifetime of the $\text{D}_{5/2}$ state in $\text{Ca}^{+}$, so spontaneous decay contributes $<0.1\%$ to the infidelity. Spontaneous decay can also occur during the measurement protocol, but that is included in the reported error from state preparation and measurement.

\section{Theory}\label{sec:theory}

\subsection{Derivation of pulse sequences using quantum signal processing}\label{sec:derivation}

With access to multiple oracle queries, we can construct algorithms that accurately and unambiguously perform quantum channel discrimination for certain classes of channels using the techniques of quantum signal processing \cite{martyn_grand_2021}. Note that in this section, variables called $\theta$ refer to the angles of processing pulses, and variables called $\phi$ refer to the oracle signal angle, regardless of if we are discussing the ASK or PSK cases, keeping with the notation for QSP from \cite{martyn_grand_2021}.

\subsubsection{Three-angle oracles}

Following \cite{rossi_quantum_2021}, we construct an algorithm that determines which of the three signal angles ($\phi_0=0$, $\phi_1=2\pi/3$, or $\phi_2=4\pi/3$) was applied by the oracle with two steps, each providing one bit of information about the answer. The first step tells us whether or not the signal angle was 0, and if it was not, the second step tells us whether the signal angle was $2\pi/3$ or $4\pi/3$. For the first step, we want the algorithm to perform $U_\mathrm{QSP}(0) = \mathbbm{1}$ if the angle was 0. If the angle was $2\pi/3$ or $4\pi/3$, we want the algorithm to do a bit flip, such as $U_\mathrm{QSP}(2\pi/3) = U_\mathrm{QSP}(4\pi/3) = X$, where $X=\sigma_x$ is the SU(2) quantum \textsc{not} gate. To construct the second half of the algorithm, we can simply apply an extra rotation before each application of the oracle, mapping the oracle rotations such that the algorithm performs the identity for $\phi = 2\pi/3$, and a bit flip for the other two angles. In the experiment, these two steps are combined and performed in the higher-dimensional Hilbert space of our ions, with measurement deferred until the end of the experiment. Thus, this multi-step protocol can be realized experimentally as a single-shot algorithm.

The formalism of QSP can be described using a parameterized signal rotation $\mathcal{W}(a)$ (which we call the oracle) of the form
\begin{equation}
    \mathcal{W}(a) = \left[
    \begin{matrix}
    a & i\sqrt{1-a^2} \\
    i\sqrt{1-a^2} & a
    \end{matrix} \right], 
    \label{qsp-signal}
\end{equation}
where $a = \cos{(\phi/2)}$. Substituting $\cos(\phi/2)$ for $a$ makes it clear that $\mathcal{W}(a)$ is simply a rotation about the $\hat{x}$ axis of the Bloch sphere by angle $\phi$. The quantum signal processing theorem \cite{martyn_grand_2021} states that interleaving $z$-rotations by angles $\vec{\theta} = \{\theta_0, \dots, \theta_d\}$ with the oracle generates a $d$-degree polynomial transformation of $a$ in the top left element of the matrix:
\begin{eqnarray}
    U_{\vec{\theta}}(a) &=& e^{i \theta_0 \sigma_z} \prod\limits_{k=1}^d \mathcal{W}(a) e^{i \theta_k \sigma_z} \\
     &=&
    \begin{bmatrix}
    P(a) & iQ(a)\sqrt{1-a^2} \\
    iQ^*(a)\sqrt{1-a^2} & P^*(a)
    \end{bmatrix},
    \label{eq:qsp-poly}
\end{eqnarray}
where $\sigma_z$ is the Pauli $z$ matrix, $P(a)$ is a (complex-valued) $d$-degree polynomial in $a$, and $Q(a)$ is another polynomial in $a$ such that the resulting operation is unitary. The possible polynomial transformations $P(a)$ are limited only by unitarity, parity, and degree (see Thm. 1 of \cite{martyn_grand_2021} for details). In our experiments, we have access to $x$ and $y$ rotations of the Bloch sphere by setting the phase of the applied rf radiation, but a simple remapping of $\hat{x} \mapsto \hat{y}$, $\hat{y} \mapsto \hat{z}$, and $\hat{z} \mapsto \hat{x}$ allows us to use the pulse sequences generated with QSP. The QSP theorem also states that, for any polynomial $P(a)$ meeting the requirements of the theorem, it is always possible to find rotation angles $\vec{\theta}$ (known as ``quantum signal processing phases'') that lead to the desired polynomial in the upper-left element of the resulting matrix $U_{\vec{\theta}}(a)$. There exist algorithms that stably and efficiently find these QSP phases, such as \textsc{pyqsp} \cite{chao_finding_2020, martyn_grand_2021}.

As discussed in \cite{rossi_quantum_2021}, for a set $\{a_i\}$ of interest, we can choose polynomials $P(a)$ that transform the signal (oracle) rotations $W(a_i)$ such that $\left| \bra{0} U_{\vec{\theta}}(a_i) \ket{0} \right|^2$ is 1 for the particular $a_i$ we choose and 0 for all of the other members of $\left\{a_i\right\}$. For three signals, we can first distinguish between $a_0 \mapsto 1$ and $\{a_1, a_2\} \mapsto 0$. A modified polynomial can then be used to distinguish $a_1$ from $a_2$ in the second step of the algorithm.

In order to discriminate between $\{0\}$ and $\{2\pi/3, 4\pi/3\}$, we may use the degree-$2$ polynomial,
\begin{equation}
    P(a) = \frac{4}{3}a^2 - \frac{1}{3} ,
    \label{eq:bisecting-poly}
\end{equation}
that maps $P(\cos{0}) \mapsto 1$ and $P(\cos{2\pi l/3}) \mapsto 0$ for $l=1,2$. Thus, measuring in the computational basis, we distinguish with certainty between $\phi_0$ and $\{\phi_1, \phi_2\}$. We can then distinguish between the remaining possibilities by appending an extra $x$-rotation of angle $2\pi/3$ to each signal rotation call. This cycles the signal rotations corresponding to different phase parameters $(\phi_0, \phi_1, \phi_2) \mapsto (\phi_2, \phi_0, \phi_1)$, and using the same polynomial \cref{eq:bisecting-poly} on this modified oracle distinguishes between $\phi_1$ and $\phi_2$.

\subsubsection{Scaling to larger sets of channels}

Scaling to large numbers of phases is straightforward. For a set of $2^k = n$ equally spaced phase angles on the unit circle beginning from $\phi_0 = 0$, we may use a bisecting procedure \cite{rossi_quantum_2021} to rule out half of the remaining angles with each measurement. Given the convenient symmetry of $2^k$ phases spaced equally on the unit circle, the Chebyshev polynomials of the first kind $T_d(\cos{\theta}) = \cos{d \theta}$ will successively bisect the search space and require in total $2n$ oracle queries to determine the phase with certainty.

To see this, assume we can distinguish the set of $n' = 2^{k-1}$ phase angles $\Phi_{k-1} = \{0, \frac{2\pi}{n'}, \dots, \frac{(n'-1)\pi}{n'}\}$. Then, we want to show that we can distinguish between two subsets of the next case for $n = 2^k$. We can separate $\Phi_k$ into two distinct subsets given by $\Phi_{k-1}$ and $\Phi_{k-1} + \frac{2\pi}{n}$. Because the Chebyshev polynomial of degree $n/4$, $T_{n/4} (\cos \theta) = \cos{\frac{n}{4}\theta}$, has zeros at $\theta \in \Phi_{k-1}$ and is 1 at $\theta \in \Phi_k$, this polynomial will distinguish with certainty between the two subsets. Note that because the two subsets we distinguish between are related by a constant offset, the protocol for distinguishing $\Phi_{k-1}$ angles will work for both subsets as the additional factor of $2\pi/n$ in one subset will not affect the placement of the nodes and peaks on the lower degree Chebyshev polynomials. The base case of $n=2$ (where the oracle is either $I$ or $\sigma_x$) is trivial to distinguish, by simply applying the oracle once and measuring.

For other numbers of channels $n$ that are not powers of 2, protocols for each of the prime factors of $n$ can be concatenated in a similar manner. These are described in \cite{rossi_quantum_2021}, which shows than any arbitrary number of channels $n$ can be distinguished in $O(n)$ queries.

\subsection{PSK channel discrimination}\label{sec:psk}

While we would also like to be able to discriminate phase-encoded signals, the algorithms presented in \cite{rossi_quantum_2021} do not include protocols for distinguishing the quantum analogue of PSK signals, which are $\pi$-rotations with differing phases. Here, we show that it is possible to add additional rotation operators before and after this sort of PSK oracle such that the resulting unitary operations are closely related to the ASK oracles discussed previously. Thus, we can extend the schemes in \cite{rossi_quantum_2021} to allow for the discrimination of PSK signals. We are looking for additional processing pulses that can be added before and after a PSK oracle that convert the oracle's $\pi$-rotations about different axes into rotations about the $\hat{x}$-axis of different rotation angles. Consider the phase oracle $R(\pi, \phi) = e^{-i \frac{\pi}{2} \sigma_{\phi}}$. This is a $\pi$-rotation on the Bloch sphere about axis $\hat{x} \cos\phi + \hat{y} \sin\phi$. However, this can be converted into the desired form by adding $y$ and $z$ rotations:
\begin{equation}
    R_z(\pi) R_y(-\pi/2) R(\pi, \phi) R_y(\pi/2) = R_x(2\phi),
\end{equation}
where $R_i(\theta) = e^{i \frac{\theta}{2} \sigma_i}$ for $i \in \{x, y, z\}$.

To then perform PSK channel discrimination, there is an additional issue to address. Converting the phase of the pulse to a rotation angle takes the rotation from $R(\pi, \phi) \mapsto R_x(2\phi)$, i.e. it takes a $\pi$-pulse about the axis at angle $\phi$ in the $x$-$y$ plane of the Bloch sphere to an $x$ rotation of angle $2\phi$. This can be seen in Fig. 4 of the main text, as the phase-encoded protocol in panel (c) has extra peaks at $\pi/3$, $\pi$, and $5\pi/3$ compared to the angle-encoded case in panel (b). Because $\phi$ can be from 0 to $2\pi$, the values of the encoded angle $2\phi$ now range from $0$ to $4\pi$. Because $R_x(\phi + 2\pi) = -R_x(\phi)$, standard QSP can only distinguish angles modulo $2\pi$ because the only difference is a global phase. Clearly, there will be ambiguity for any case where the number of phases is even. In this case, one can use the standard QSP algorithm up to this ambiguity, at which point there will be two possible signal angles, $\phi'$ and $\phi'+\pi$ for $\phi' \in [0, \pi)$.

To distinguish these angles, the use of auxiliary states is required to turn the global phase into a relative phase. Creating an equal superposition of two states with a Hadamard gate, applying the transformed oracle $R_x(2\phi)= \pm R_x(2\phi')$ to both subspaces, and then applying the Hadamard again allows the $\pm 1$ phase to be directly measured, resolving the ambiguity. This means, for example, that the three-angle PSK scheme described in the main text becomes a six-angle PSK scheme with the addition of one extra oracle query. Because we are already using the larger Hilbert spaces of trapped ions, this is not a burdensome requirement.

\subsection{Query complexity and quantum advantage} \label{sec:queries}
Due to the ability to query quantum channels multiple times without disturbing the channel itself, it is possible to discriminate a set of channels with certainty using a finite number of queries. This is in contrast to quantum process tomography, which allows measurement of arbitrary quantum channels, but requires infinite queries of the process to achieve perfect discrimination \cite{nielsen_quantum_2000}.  There are a variety of optimal strategies that achieve perfect discrimination with either entanglement \cite{acin_statistical_2001, dariano_using_2001} or multiple queries of the channel \cite{duan_entanglement_2007, duan_perfect_2009}. On the other hand, if only a single entanglement-free query of the channel is permitted, the problem reduces to that of quantum \textit{state} discrimination and is limited by the Helstrom bound \cite{herzog_minimum-error_2002, chefles_unambiguous_1998, bae_quantum_2015}. Such a single-query protocol consists of preparing an initial state of one's choosing, sending that state through the quantum channel, and measuring in any chosen basis.

It has been shown previously that two angles can be discriminated perfectly by using multiple oracle queries. This could be extended to an arbitrary number of angles by adapting these schemes to eliminate one angle at a time. However, the protocol in \cite{rossi_quantum_2021} further reduces the number of queries by using a binary search to rapidly eliminate possible signal angles. This reduces the query complexity for distinguishing between $n$ symmetric angles from $O(n^2)$ (as each comparison between a pair of angles requires $O(n)$ queries) to $O(n)$, a quadratic speedup compared to repeated use of earlier two-unitary discrimination protocols \cite{rossi_quantum_2021}.

Additionally, we demonstrate an advantage over simple incoherent protocols, even including experimental imperfections. The best probability of guessing the correct state with a single query and measurement is $p_\text{guess} = 2/3$ for these three symmetric states \cite{ban1997optimum, bae_quantum_2015, solis-prosser_experimental_2017}, saturating the Helstrom bound. This measurement scheme is also known as a minimum-error (ME) measurement scheme. An incoherenct ME measurement scheme with four oracle queries decided with a majority vote of the four measurement outcomes only has an approximately 74\% chance of correctly determining the oracle value, significantly lower than the ${\tt>}99\%$ accuracy we report for our experiments where we coherently query the channel. Even in the case that all four of the ME measurements give the same results, the case where the incoherent protocol has the highest confidence, the probability of guessing correctly is still only $p_\text{guess} \approx 98.8\%$, so the confidence of our coherent protocol is always higher than any four ME measurements.

We can show the same for the case of unambiguous discrimination (UD). The optimal probability of unambiguously discriminating which of three symmetric states is present, from \cite{bergou2012optimal}, is
\begin{equation}
    P_\mathrm{UD} = \eta_1 \frac{s_{12} s_{13}}{s_{23}} + \eta_2 \frac{s_{12} s_{23}}{s_{13}} + \eta_3 \frac{s_{13} s_{23}}{s_{12}},
\end{equation}
where $\eta_i$ is the \textit{a priori} probability of state $i$ being sent, and $s_{ij} = \left|\braket{\psi_i|\psi_j}\right|$ is the magnitude of the overlap between states $i$ and $j$. For three equally-probable symmetric states (as produced by the oracles we consider acting on $\ket{0}$), $\eta_1 = \eta_2 = \eta_3 = 1/3$, and $s_{12}=s_{23}=s_{13} = 1/2$. Thus, the probability of an optimal UD scheme returning a definitive answer (as opposed to the inconclusive result) is $P_\mathrm{UD} = 1/2$. Therefore, four trials of this unambiguous discrimination scheme have a maximum probability of correctly identifying the state of $1-(1/2)^4 \approx 94\%$. 

Thus, for both of the well-understood limiting cases of quadruple ME and UD incoherent measurements, we have shown that our coherent QSP-based channel discrimination scheme determines the oracle correctly with higher probability even when including experimental errors, demonstrating a quantum advantage. While this does not preclude the possibility of a measurement scheme in the continuum between ME and UD measurement from having a higher success probability, or of some entirely different adaptive measurement scheme, no such techniques are known. In the optimal case of our coherent QSP protocol, of course, the error goes to 0, and the ${\tt<}1\%$ errors we report are due only to experimental imperfections and not an intrinsic limit on the accuracy of the protocol.

\subsection{Applicability of QSP to the \texorpdfstring{$\bm{\text{D}_{5/2}}$}{D\_5/2} manifold} \label{sec:su6}

Running the algorithm in the combined $\text{D}_{5/2}$ and $\text{S}_{1/2}$ state space of an ion allows us to perform single-shot measurement of a set of several quantum channels. Making use of the full $\text{D}_{5/2}$ manifold for information processing (rather than isolating a qubit as described in \cref{sec:magnetic} for the magnetic field compensation routine) also has experimental advantages. The Rabi frequency of the rf operations is not limited by the achievable magnitude of the light shift (in the magnetic field compensation procedure, we apply rf radiation with a $\pi$-time of approximately \SI{350}{\micro s}, compared to the $\sim$\SI{50}{\micro s} $\pi$-time of the rf operations used in the channel discrimination algorithm). Additionally, selection rules do not forbid the light-shifting \SI{729}{nm} laser from also applying a small shift to the upper qubit state ($m=-3/2$), a shift which fluctuates with laser amplitude fluctuations. The longer duration of a light-shifted version of the algorithm would suffer from even more stringent frequency stability requirements as well as increased spontaneous emission errors from the $\text{D}_{5/2}$ manifold. Prior experiments \cite{sherman_experimental_2013, bazavan_synthesizing_2022} have used the $\SI{854}{nm}$ transition to the $\text{P}_{3/2}$ state to light shift states out of resonance and create a qubit, but this relies on having pure circular polarization to avoid pumping population out of the qubit manifold.

The algorithm proposed in \cite{rossi_quantum_2021} was originally a multi-step protocol using multiple measurements to provide each bit of information about the oracle's signal angle, but it is straightforward to adapt it to a single-shot protocol in larger state space. By using the many ground and metastable states of an ion (as opposed to just a qubit), measurements of each bit of information can be delayed until after both halves of the algorithm have run by shelving the results. For our implementation of these algorithms, we shelve the result of the first step of the protocol to a state in the $\text{S}_{1/2}$ manifold, and then continue with the second half of the protocol in the $\text{D}_{5/2}$ manifold.

This implementation of the PSK algorithm described in the main text was developed using numerical optimization techniques, but this is not required. Our implementation takes advantage of the structure of the spin-\sfrac{5}{2} system to use fewer processing pulses than a direct application of quantum signal processing would. However, this is not necessary, and the ASK algorithm we implemented was a direct implementation of the algorithm described in \cite{rossi_quantum_2021}, with QSP phases found with \textsc{pyqsp}.

\begin{figure}[t]
    \centering
    \includegraphics[width=0.48\textwidth]{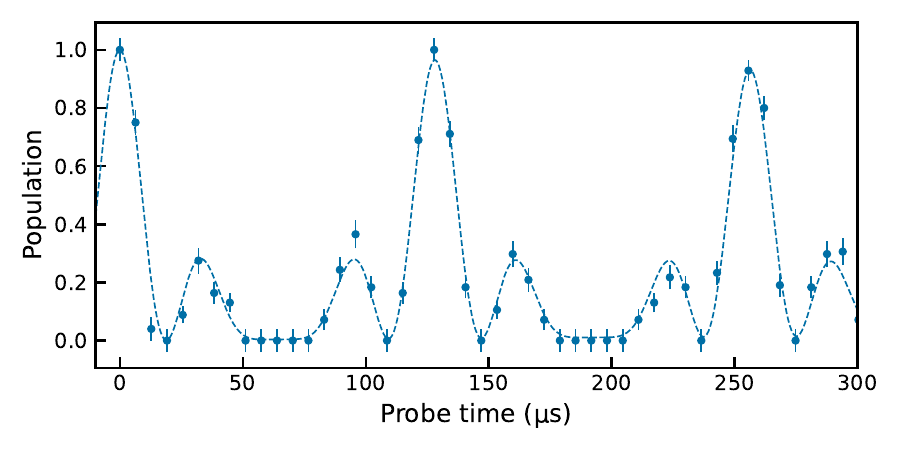}
    \caption{Generalized Rabi oscillations in the six-level $\text{D}_{5/2}$ manifold, with a $\pi$ pulse taking approximately \SI{55}{\micro\second}. Data points are overlaid on a fit to an analytic solution for the six-level Rabi oscillations shown as a dashed line. State preparation is performed by the \SI{729}{nm} laser, initializing to the $m=-3/2$ magnetic sublevel of the metastable manifold, as shown in \cref{fig:su6_isolation}(a).}
    \label{fig:su6_rot}
\end{figure}

We can use pulse sequences for this decision problem developed with standard spin-\sfrac{1}{2} QSP in the larger $\text{D}_{5/2}$ state space because the spin-\sfrac{5}{2} angular momentum operators behave very similarly to their spin-\sfrac{1}{2} counterparts \cite{curtis2010measurement}. A thorough description of the dynamics of spin-$J$ systems (for arbitrary $J$) acted on by resonant radiation can be found in \cite{cook1979coherent}. At first, our ability to implement algorithms in the $\text{D}_{5/2}$ manifold may seems surprising, as the operators at our disposal ($J_x^{(6)}$ and $J_y^{(6)}$) are not sufficient for arbitrary control of the $\text{D}_{5/2}$ 6-level system \cite{giorda_universal_2003}. This means that, in general, we cannot apply arbitrary SU(6) matrices of our choosing. However, due to the commutation relations of the SU(6) angular momentum operators, we can still implement certain algorithms. For any set of eigenstates defined by a quantum spin $\vec{J}$, regardless of the size of the state space $2J+1$, we can write the Hamiltonian applied by the rf signal (in the rotating frame) as $H_\mathrm{I} = \vec{\mu} \cdot \vec{B} = g_J \mu_\mathrm{B} B [J_x \cos(\phi) + J_y \sin(\phi)]$, where $g_J$ is the Landé $g$-factor, $\mu_\mathrm{B}$ is the Bohr magneton, $B$ is the amplitude of the applied oscillating magnetic field, $\phi$ is the phase of the applied field in the lab frame, and $J_x$ and $J_y$ are the appropriate spin-\sfrac{1}{2} or spin-\sfrac{5}{2} angular momentum operators. The spin-\sfrac{5}{2} angular momentum $J_i^{(6)}$ operators obey the same commutation relations as in spin-\sfrac{1}{2}: $[J_i, J_j] = i \hbar \epsilon_{ijk} J_k$, where $\epsilon_{ijk}$ is the antisymmetric Levi-Civita symbol. Thus, the algebra of rotation operators generated by our Hamiltonian behaves the same in all cases, and we can still perform many interesting quantum signal processing algorithms in the $\text{D}_{5/2}$ state space.

To see how this allows us to implement certain algorithms, consider our example of channel discrimination (a decision problem). The algorithm developed in \cite{rossi_quantum_2021} results in the application of either the identity or a $\pi$-rotation (bit flip) operator, and simply converting the SU(2) rotation sequence to SU(6) preserves these two outcomes. This conversion is simply the replacement of the spin-\sfrac{1}{2} angular momentum operators $J_i^{(2)} = \sigma_i/2$ with the spin-\sfrac{5}{2} versions $J_i^{(6)}$. Due to the identical commutation relations of these operators, sequences of rotations that add up to either the identity or a bit flip operation in SU(2) likewise add up to the SU(6) generalization of these operations:
\begin{align}
X^{(2)} &= -i e^{i \pi J_x^{(2)}} \\
&= \begin{bmatrix}
    0 & 1 \\
    1 & 0
\end{bmatrix}, \nonumber
\end{align}
\begin{align}
X^{(6)} &= -i e^{i \pi J_x^{(6)}} \\
&= \begin{bmatrix}
    0 & 0 & 0 & 0 & 0 & 1 \\
    0 & 0 & 0 & 0 & 1 & 0 \\
    0 & 0 & 0 & 1 & 0 & 0 \\
    0 & 0 & 1 & 0 & 0 & 0 \\
    0 & 1 & 0 & 0 & 0 & 0 \\
    1 & 0 & 0 & 0 & 0 & 0
\end{bmatrix}.\nonumber
\end{align}
Here, the two-dimensional matrix acts on the standard computational basis states for a qubit. For the SU(6) operator $X^{(6)}$, we take the basis states to be the magnetic sublevels of the $\text{D}_{5/2}$ manifold, $\left\{\ket{m_i}\right\}$ for $m_i \in -5/2, -3/2, \dots, 5/2$. Thus, any population initialized in the $m=m_i$ magnetic sublevel of the $\text{D}_{5/2}$ manifold would end up in the $m=-m_i$ magnetic sublevel, much like the results of the  two-level \textsc{not} gate $X^{(2)}$. Rabi oscillations in the $\text{D}_{5/2}$ manifold are shown in \cref{fig:su6_rot}, where in this case we prepare the $m=-3/2$ state as shown in \cref{fig:su6_isolation}(a). The rotations are described by $U(t) = e^{i \Omega t J_x^{(6)}}$, where $\Omega$ is the Rabi frequency and $t$ is the probe time, such that $\theta = \Omega t$. We can then define a time $t_\pi = \pi/\Omega$. Applying $U(t_\pi)$ results in  $X^{(6)}$ being applied, and all population has been moved from the $m=-3/2$ state to the $m=+3/2$ state, so after de-shelving of the $m=-3/2$ state the ion remains dark. After $2t_\pi$, the identity has been applied up to a global phase, and population returns to the $m=-3/2$ state.

%% file: main.bbl
\begin{thebibliography}{69}
\bibitem [{wil()}]{wilson_footnote}%
  \BibitemOpen
  \href@noop {} {\bibinfo {title} {The {Wilson} score interval is recommended
  by statisticians over the commonly used {Wald} interval}}\BibitemShut
  {NoStop}%
\end{thebibliography}
